\documentclass[accepted]{uai2023} 

\usepackage[american]{babel}

\usepackage{natbib} 
\usepackage{mathtools} 
\usepackage{booktabs} 
\usepackage{tikz} 
\usepackage{times}
\usepackage{soul}
\usepackage{url}
\usepackage{hyperref}
\hypersetup{
    colorlinks,
    linkcolor=blue,
    citecolor=blue,
    urlcolor=blue
}
\usepackage[utf8]{inputenc}
\usepackage{graphicx}
\usepackage{amsfonts}
\usepackage{booktabs}
\usepackage{subfigure}
\usepackage{amssymb}
\usepackage{xspace}
\usepackage{textcase}
\usepackage{amsmath}
\usepackage{amsthm}
\usepackage{mathtools}
\usepackage{enumitem}
\usepackage{refcount}
\usepackage{booktabs}
\usepackage[ruled,vlined,linesnumbered]{algorithm2e}
\usepackage{mathrsfs} 
\usepackage[font=footnotesize,labelfont=bf]{caption}
\usepackage{xcolor}
    \definecolor{darkgreen}{rgb}{0.0, 0.2, 0.13}
    \definecolor{cadmiumgreen}{rgb}{0.0, 0.42, 0.24}
    \definecolor{byzantium}{rgb}{0.44, 0.16, 0.39}
\usepackage{newfloat}
\usepackage{chngcntr}

\DeclareFloatingEnvironment[name=Table]{tablefigure} 

\makeatletter\let\c@tablefigure\c@table\makeatother 

\makeatletter\let\ftype@tablefigure\ftype@table\makeatother

\mathchardef\mhyphen="2D 

\SetCommentSty{commentstyle}
\SetKwInOut{Input}{input}
\SetKwInOut{Output}{output}

\newtheorem{theorem}{Theorem}[section]

\newtheorem{definition}{Definition}[section]

\providecommand{\super}[1]{\textsuperscript{#1}}

\providecommand{\bs}[1]{\boldsymbol{#1}}

\providecommand{\KstarMAP}{\NoCaseChange{K\super{*}MAP}\xspace}
\providecommand{\Kstar}{\NoCaseChange{K\super{*}}\xspace}

\providecommand{\AOBBKstar}{\NoCaseChange{AOBB-}\Kstar}
\providecommand{\BBKstar}{\NoCaseChange{BB}\Kstar}
\providecommand{\wMBEKstar}{\NoCaseChange{wMBE-}\Kstar}

\providecommand{\X}{\ensuremath{\boldsymbol{X}}}
\providecommand{\xx}{\ensuremath{\boldsymbol{x}}}

\providecommand{\M}{\ensuremath{\mathcal{M}} }
\providecommand{\Mcpd}{\ensuremath{\mathcal{M}_{cpd}}}

\providecommand{\D}{\ensuremath{\boldsymbol{D}}}
\providecommand{\F}{\ensuremath{\boldsymbol{F}}}

\providecommand{\vphi}{\ensuremath{\boldsymbol{\varphi}}}
\providecommand{\R}{\ensuremath{\boldsymbol{R}}}
\providecommand{\rr}{\ensuremath{\boldsymbol{r}}}
\providecommand{\E}{\ensuremath{\boldsymbol{E}}}

\providecommand{\C}{\ensuremath{\boldsymbol{C}}}
\providecommand{\CC}{\ensuremath{\boldsymbol{\mathscr{C}}}}
\providecommand{\Y}{\ensuremath{\boldsymbol{Y}}}

\providecommand{\PT}{\ensuremath{\mathcal{T}}}

\providecommand{\shrink}[1]{}

\title{ Boosting AND/OR-Based Computational Protein Design:\\
        Dynamic Heuristics and Generalizable UFO
      }

\author[1]{\href{mailto:<pezeshkb@uci.edu>?Subject=AND-OR Branch-and-Bound for Computational Protein Design Optimizing K*}{Bobak Pezeshki}{}}
\author[2]{\href{mailto:<radu.marinescu@ie.ibm.com>?Subject=AND-OR Branch-and-Bound for Computational Protein Design Optimizing K*}{Radu Marinescu}{}}
\author[1]{\href{mailto:<ihler@ics.uci.edu>?Subject=AND-OR Branch-and-Bound for Computational Protein Design Optimizing K*}{Alexander Ihler}{}}
\author[1]{\href{mailto:<dechter@ics.uci.edu>?Subject=AND-OR Branch-and-Bound for Computational Protein Design Optimizing K*}{Rina Dechter}{}}
\affil[1]{%
    University of California, Irvine
}
\affil[2]{%
    IBM Research
}

\begin{document}

\setlength{\abovedisplayskip}{3pt}
\setlength{\belowdisplayskip}{3pt}
\setlength{\textfloatsep}{6pt plus 1.0pt minus 2.0pt}
\setlength{\floatsep}{6pt plus 1.0pt minus 2.0pt}
\setlength{\intextsep}{6pt plus 1.0pt minus 2.0pt}

    \maketitle

    \begin{abstract}
        \noindent Scientific computing has experienced a surge empowered by advancements in technologies such as neural networks.  However, certain important tasks are less amenable to these technologies, benefiting from innovations to traditional inference schemes.  One such task is protein re-design.  Recently a new re-design algorithm, \AOBBKstar, was introduced and was competitive with state-of-the-art \BBKstar on small protein re-design problems. However, \AOBBKstar  did not scale well.  In this work, we focus on scaling up \AOBBKstar and introduce three new versions: \AOBBKstar-b (boosted), \AOBBKstar-DH (with dynamic heuristics), and \AOBBKstar-UFO (with underflow optimization) that  significantly enhance scalability.
    \end{abstract}

    \section{Introduction}
        Computational protein design (CPD) is the task of creating (or re-designing) proteins to achieve a desired functionality.  There are two general classes of CPD: \textit{de novo} protein design and protein re-design - the former being the creation of novel proteins and the latter the task of mutating existing proteins to enhance their functionality.
        
        Advances in \textit{de novo} protein design have been accelerated by evolving tools for protein structure prediction \citep{jumper2021alphaFold, casp15-2022-protein-structure-prediction-competition} and sequence design \citep{dauparas2022robust}, both leveraging advances in neural networks.  These methods excel in producing small binding proteins that can act as activators or inhibitors.  However, these tools often operate within a flexible framework, with loose or no constraints on the final sequence or three-dimensional structure of the designed protein.   Additionally, neural network-based protein design tools rely heavily on sequence alignment and leveraging homologies with known proteins for prediction \citep{AlLazikani2001-description-and-limitations-of-protein-structure-prediction, defresne-protein-design-with-deep-learning}.
        
        In contrast, protein \textit{\textbf{re}}-design involves mutating the amino acid sequence of a known protein structure to enhance its properties, such as stability, ligand affinity, or inhibitor resistance. A prominent software tool for comprehensive protein re-design, specifically for improving bonding affinity through optimization, is OSPREY (Open Source Protein REdesign for You) \citep{hallen18-osprey-3}. OSPREY utilizes \BBKstar, a best-first search algorithm that optimizes the \Kstar objective function \citep{ojewole18-bbkstar, hill87-statistical-mechanics-kstar, mcquarrie00-statistical-mechanics-kstar}.
        
        Recently, an alternative scheme called \AOBBKstar was introduced for protein re-design \citep{pezeshki2022uai-aobb-for-cpd-2022}. \AOBBKstar is an exact AND/OR branch-and-bound approach that incorporates a weighted mini-bucket derived heuristic. It demonstrated competitive performance compared with \BBKstar on small protein problems with rigid backbones and side-chain rotamers. However, \AOBBKstar did not scale well and had limited  effectiveness on larger problems.

        In this work, we present modifications to key components of \AOBBKstar and introduce new generalizable schemes to enhance its scalability. Our contributions are as follows:
        
        \begin{itemize}
            \item \textbf{\AOBBKstar-b (boosted):} A modified version of \AOBBKstar with a stronger \wMBEKstar{} heuristic and modifications for more efficient search.
            
            \item \textbf{\AOBBKstar-DH:} \AOBBKstar with dynamic heuristics.
            
            \item \textbf{UFO:} An approximation scheme that introduces determinism to empower constraint propagation.
            
            \item \textbf{\AOBBKstar-UFO:} UFO empowered \AOBBKstar.
            
            \item \textbf{Empirical analysis:} Evaluation of the proposed schemes on 62 real protein benchmarks comparing with previous \AOBBKstar and state-of-the-art \BBKstar.
        \end{itemize}

    \section{Background}

        We begin with some relevant background.  (An extended background is provided in the \href{https://www.ics.uci.edu/~dechter/publications.html}{Supplemental Materials}).
    
        \subsection{Computational Protein Design}
            We consider the protein \emph{re-design} task within the broader Computational Protein Design (\textbf{CPD}) field where known proteins are modified to alter their functions or interactions \citep{gainza2016algorithms}.
            Specifically, some of the amino acid positions (or \textbf{residues}) of a given protein are deemed as \textbf{mutable} - these are amino acid positions where different amino acid mutations will be considered - and a preferred sequence is determined \citep{donald11-book-algorithms-in-structural-biology}. During the process, sets of mutations are investigated, each representing a distinct amino acid sequence.  By considering a sequence (or even a partial sequence in some methods), it is possible to estimate the resulting protein's quality. This quality assessment is based on an analysis of the potential conformations of the protein's backbone and amino acid side-chains.  The conformational state space for these structures is continuous, and even when discretized remains extremely large, resulting in an intractable problem. To address this, several simplifications can be made: (i) limiting consideration to a subset of mutable residues, (ii) discretizing side-chain conformations into rotamers, and (iii) assuming a fixed protein backbone conformation. With these simplifying assumptions, numerous algorithms have been developed to identify mutations that have the potential to enhance protein functionality \citep{hallenProvableAlgsCPD, zhou16-ao-bnb-cpd}.
        
        \subsection{K\super{*} and \KstarMAP}
            Modifying the affinity between protein subunits is a common objective in protein re-design. The affinity between two protein subunits $P$ and $L$ is related to the rate at which they bind into a complexed structure $P\!L$ and dissociate back into separated $P$ and $L$ subunits (as indicated by the chemical equation: $P + L \rightleftharpoons P\!L$).  This equilibrium is described by a constant $K_a$ that can be determined \textit{in vitro} by computing the ratio of persisting concentrations of each species, $K_a = \frac{[P\!L]}{[P] [L]}$ \citep{rossotti1961stabilityconstants}. However, in order to compare $K_a$ values of various designs in this manner, it would be necessary to synthesize protein subunits through molecular processes that are both timely and costly.

            Following previous work that assumes discretization of protein structure conformations, letting 
            $E_{s(c)}$ represent the energy of a particular conformation $c$ of protein strand(s) involved in structure $s \in \{P, L, P\!L\}$, $\mathscr{R}$ the universal gas constant, $T$ temperature (in Kelvin), and $\D(C_{s})$ the discretized conformation space,
            $K_a$ can instead be approximated by
                    $
                    K^* = \frac{Z_{P\!L}}{Z_{P} Z_{L}},  \;\;\;\;
                        Z_{s} = \sum\limits_{c \in \D(C_{s})} e^{ -\frac{E_{s(c)}}{\mathscr{R} T} } 
                    $
            \citep{lilien04-protein-redesign}. Due to the independence within different states of the protein, we can generalize further as: $K^* = \frac{Z_{B}}{Z_{U}}$, where $B$ represents the bound (complexed) structure(s) and $U$ represents the unbound (dissociate) structures. For a two-subunit system, $B = \{P\!L\}$ and $U = \{P\} \cup \{L$\}. This generalized representation can also be used for K\super{*} computations involving more than two subunits.
                    
            A common goal in protein redesign is to maximize protein-ligand interaction such as by finding an amino acid assignment $\R\!\leftarrow\!\rr$ that maximizes \Kstar:
            \begin{align}
                K^{*}\textnormal{MAP} = \max_{\R} K^*(\rr)
            \end{align}
            
            \begin{figure}[!tb]
                \centering
                \includegraphics[scale=0.29]{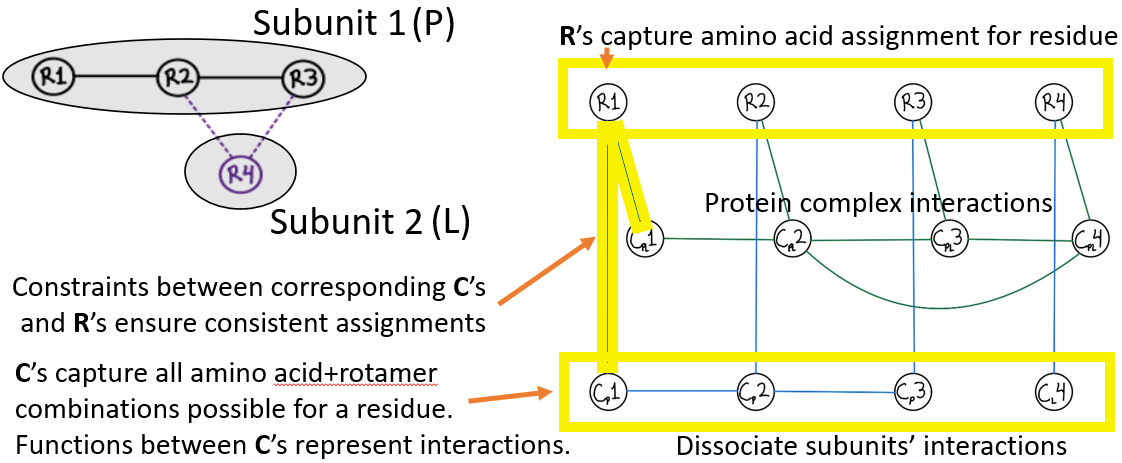}
                \caption{CPD formulated as a graphical model.}
                \label{fig:cpd-pgm}
            \end{figure}

        \vspace{-8pt}
        \subsection{Graphical Model for \KstarMAP}

            A discrete graphical model can be defined as a 3-tuple $\mathcal{M} \! = \! \langle \X,\D,\F \rangle$, where: $\mathbf{X}$ is a set of variables for which the model is defined; $\mathbf{D} \! = \! \{D_{X} \! : X \! \in \! \mathbf{X}\}$ is a set of finite domains, each defining the possible assignments for a variable; each $f_{\alpha} \in   \mathbf{F}$ is a real-valued function defined over a subset of the model's variables $\alpha \subseteq \mathbf{X}$ known as the function's \textbf{scope}.  More concretely, if we let $D_{\alpha}$ denote the Cartesian product of the domains of the variables in $\alpha$, then $f_{\alpha} \! : D_{\alpha} \! \rightarrow \mathbb{R}_{\geq 0}$. We let capital letters ($X$) represent variables and small letters ($x$) represent their assignment. Boldfaced capital letters ({\bf X}) denote a collection of variables, $|{\bf X}|$ its cardinality, $D_{\X}$ its joint domain, and $\xx$ a particular realization in that joint domain called a \textbf{configuration}. Operations denoted $\bigoplus_{\X}$ (ex. $\sum_{\X}$) imply
            $\bigoplus_{\X} \implies \bigoplus_{\xx \in D_{\X}}$.
                
            An important graphical model task that resembles \Kstar optimization is determination of the \textbf{marginal maximum a-posteriori (or MMAP)} (Definition \ref{def:mmap},  \citep{dechter-book-2ndEd}). Given the similarity, many graphical model algorithms for MMAP can be leveraged for \Kstar optimization.
            
            \begin{definition}[MMAP] \label{def:mmap}
            Given a graphical model $\mathcal{M} \! = \! \langle \X,\D,\F \rangle$, the marginal maximum a-posteriori of $\M$ is:
                \begin{align}
                    MMAP(\mathcal{M}, \bs{Q} \subset \X) = \max_{\bs{Q}} \sum_{\bs{S} = \X \setminus \bs{Q}} \prod_{\F} f(\bs{q},\bs{s})
                \end{align}                
            \end{definition}
            
            Inspired by \citet{viricel18-cfn-cpd} and \citet{vucinic19-positive-multistate-design}, a recent formulation of CPD as graphical models was proposed by \cite{pezeshki2022uai-aobb-for-cpd-2022} as described below:
                  
            \textbf{Variables and Domains:~}
                The variables of the model consist of the union of two disjoint, but related, sets: a set of variables $\R$ representing the protein residue positions and a set of conformation variables $\C$ representing spatial orientation of the amino acids at each residue. More specifically, there are residue variables $\boldsymbol{R} = \{R_i \:|\: i \in \{1,2,...,N\}\}$, one for each of the $N$ protein residue being considered during the redesign. The respective domain of each $R_{i}$ consists of the amino acids being considered for its corresponding residue position. Each $R_{i}$ is accompanied by respective conformation variables $C_{\gamma(i)}$ representing the spatial conformation of the amino acid assigned to residue $R_{i}$ when the protein is in state $\gamma \in \vphi = \{B, U\}$. Each $C_{\gamma(i)}$ has a domain $D_{C_{\gamma(i)}} = \{c \: | \: c$ is a rotamer for one of the possible amino acids of $R_i\}$.  Thus, the amino acid assignment to $R_i$ acts as a selector into the possible assignments to $C_{\gamma(i)}$.
            
            \textbf{Functions:~}
                Constraints $\CC = \{\mathscr{C}_{\gamma(i)}(R_{i},C_{\gamma(i)}) \:|$ $i \in \{1,2,...,N\}, \gamma \in \varphi \: \}$ ensure that the assigned rotamer to $C_{\gamma(i)}$ belongs to the amino acid assigned to $R_i$. Energy functions $E_{\gamma}^{sb} = \{E_{\gamma(i)}^{sb}(C_{\gamma(i)}) \:|\: i \in \{1,2,...,N\}\}$ capture the energies of interaction of the amino acid at each residue $i$ with itself and the surrounding backbone.  $E_{\gamma}^{pw} = \{E_{\gamma(ij)}^{pw}(C_{\gamma(i)},C_{\gamma(j)}) \:|$ for $i,j$ s.t.\! $R_i$ and $R_j$ interact$\}$ capture pair-wise energies of interaction.
                
            \textbf{Objective:~}\hfill\\
                \begin{small}
                    Let...
                    \vspace{-18pt}
                    \begin{align} \label{eq:boltzman-transformmed-energies-and-constraints}
                        \begin{split}
                            Z_{\gamma}(\rr) = 
                            \sum_{\C_{\gamma}} 
                            &\prod_{\bs{\mathscr{C}_{\gamma}}
                            } \mathscr{C}_{\gamma(i)}(r_i,c_{\gamma(i)})
                            \: \\[-6pt]
                            \cdot 
                            &\prod_{\bs{E_{\gamma}^{sb}}
                            } e^{-\frac{E_{\gamma(i)}^{sb}(c_{\gamma(i)})}{\mathscr{R} T}}
                            \: \\[-6pt]
                            \cdot 
                            &\prod_{\bs{E_{\gamma}^{pw}} 
                            } e^{-\frac{E_{\gamma(ij)}^{pw}(c_{\gamma(i)},c_{\gamma(j)})}{\mathscr{R} T}}
                        \end{split}
                    \end{align}
                    \vspace{-12pt}
                    Objective Function: 
                    \vspace{-11pt}
                    \begin{align}
                        K^*(\rr) &= \frac{Z_{B}(\rr)}{Z_{U}(\rr)}, \hfill
                    \end{align}
                    \vspace{-8pt}
                    Task:
                    \vspace{-11pt}
                    \begin{align}
                        K^{*}\textnormal{MAP} &= \max_{\R} K^*(\rr), \hfill
                    \end{align}
                \end{small}
                \vspace{-14pt}
      
        \subsection{\AOBBKstar Algorithm}
            \AOBBKstar is an exact branch-and-bound algorithm proposed by \citet{pezeshki2022uai-aobb-for-cpd-2022} for optimization of \Kstar.  \AOBBKstar constrains solutions such that the partition function of each protein subunit, $Z_{\gamma}$, is greater than a biologically-relevant subunit-stability threshold $S_{\gamma}$ as defined in \citep{ojewole18-bbkstar}.  The algorithm takes a CPD graphical model as input (Definition \ref{def:cpd-graph-mod}) and outputs the \KstarMAP value of an optimal amino acid assignment to the residues.  The algorithm is based on a class of AND/OR search algorithms over graphical models for optimization and  inference tasks \citep{marinescu14-aobb-mmap} and is empowered by constraint propagation.  
    
            \begin{definition}[CPD Graphical Model] \label{def:cpd-graph-mod}
                Let $\mathcal{M}_{cpd} = \langle \X = \R \cup \C,\; \D,\; \F = \CC \cup \E, \; \bs{S} \rangle$ be a CPD graphical model for \KstarMAP optimization.
            \end{definition}

            \AOBBKstar{} (Algorithm \ref{alg:AOBB-KstarMAP-MAIN}) traverses an underlying AND/OR search tree expanding nodes in a depth-first manner (line \ref{alg:AOBB-KstarMAP-MAIN:expand}), and pruning whenever any of three conditions are triggered: (1) constraint propagation finds that the current assignments are inconsistent (line \ref{alg:AOBB-KstarMAP-MAIN:CPP}), (2) a subunit-stability constraint is violated (line \ref{alg:AOBB-KstarMAP-MAIN:SSP}), or (3) it can be asserted that the current amino acid configuration cannot produce a \Kstar{} better than any previously found (line \ref{alg:AOBB-KstarMAP-MAIN:UBP}).  Backtracking occurs when all of a leaf node's children have been explored and returned from  (line \ref{alg:AOBB-KstarMAP-MAIN:backtrack}), at which point the \Kstar{} value of the sub problem the node roots is known exactly and bounds of its parents are tightened accordingly. The algorithm progresses in this manner until it finally returns to, and updates, the root of the search tree with the maximal \Kstar{} value corresponding to an amino acid configuration that also satisfies the subunit-stability thresholds.

            The version of \AOBBKstar used by \citet{pezeshki2022uai-aobb-for-cpd-2022} is guided by \wMBEKstar, a mini-bucket-based heuristic \citep{dechter99c} with cost-shifting \citep{liu11-bounding-partition-function, IhlerFDO12}, adapted for \KstarMAP. \wMBEKstar{} is described in Algorithm \ref{alg:wMBE-KstarMAP} and operates similarly to its corresponding scheme for MMAP, wMBE-MMAP \citep{marinescu14-aobb-mmap}. \wMBEKstar takes a variable ordering that constrains maximizations to be processed last (line \ref{alg:wMBE-KstarMAP:reverse-ordering}). Functions are partitioned into the buckets $B_{k}$, and for any bucket $k$ with width larger than a provided i-bound (ie. when the number of distinct variables in the bucket's functions is greater than the provided i-bound), a bounded approximation is made by partitioning the bucket functions into $T$ mini-buckets $MB_{k}^{(t)}$ (line \ref{alg:wMBE-KstarMAP:partition-mini-buckets}) and taking a power-sum over the bucket variable (lines \ref{alg:wMBE-KstarMAP:BEGIN-upper-bound-bucket}-\ref{alg:wMBE-KstarMAP:END-upper-bound-bucket}, \ref{alg:wMBE-KstarMAP:BEGIN-lower-bound-bucket}-\ref{alg:wMBE-KstarMAP:END-lower-bound-bucket}, Definitions \ref{def:consolidated-mini-bucket-function}-\ref{def:mini-bucekt-power-sum}), leveraging Holder's Inequality \citep{hardy88-inequalities}.

            \begin{definition}[Consolidated Mini-Bucket Function] \label{def:consolidated-mini-bucket-function}
                Consider a mini-bucket $MB_{k}^{(t)}$.
                We define its consolidated mini-bucket function as $f_{MB_{k}^{(t)}} := \prod_{f \in MB^{(t)}_k} f$
            \end{definition}

            \begin{definition}[Mini-Bucket Power Sum] \label{def:mini-bucekt-power-sum}
                The power sum of a mini-bucket $MB_{k}^{(t)}$ is defined as $\sum_{X}^{w} f_{MB_{k}^{(t)}} := (\sum_{X} (f_{MB_{k}^{(t)}})^{\frac{1}{w}})^{w}$.
            \end{definition}
            
            Unlike wMBE-MMAP, buckets of \wMBEKstar corresponding to variables in $\boldsymbol{C_U}$, whose marginal belongs to the denominator of the K\super{*} expression, are lower-bounded (to lead to an upper bound on K\super{*}) by using a modification to Holder's inequality that incorporates negative weights \citep{liu11-bounding-partition-function} (lines \ref{alg:wMBE-KstarMAP:BEGIN-lower-bound-bucket}-\ref{alg:wMBE-KstarMAP:END-lower-bound-bucket}). When messages are passed from buckets of $\boldsymbol{C_U}$ to that of $\boldsymbol{R}$ the messages are inverted to accommodate being part of the \Kstar denominator (line \ref{alg:wMBE-KstarMAP:invert-bucket-message}). Although details are omitted here, \wMBEKstar can also employ cost shifting to tighten its bounds (for more details see \citep{flerova2011-mbe-mm,liu11-bounding-partition-function}). 

            \paragraph{Complexity.}
                The time complexity of \AOBBKstar's search is exponential in the depth of the search tree, while the space complexity is linear in its depth. The time and space complexity of \wMBEKstar are exponential in the i-bound $i$.

            \begin{algorithm}[t!]
                \caption{\AOBBKstar{}}
                \label{alg:AOBB-KstarMAP-MAIN}
                \begin{footnotesize}
                    \SetInd{0.25em}{0.55em}
                    \DontPrintSemicolon 
                \Input{
                    CPD graphical model $\mathcal{M}_{cpd}$  (Def \ref{def:cpd-graph-mod}); \\
                    pseudo tree $\PT$ guiding node expansions;\\
                    $K^*$ upper-bounding heuristic function $h_{K^*}(.)$;\\
                    $Z_\gamma$ upper-bounding heuristic function $h_{Z_\gamma}(.)$
                }
                \Output{
                    $K^*MAP(\mathcal{M}_{cpd})$
                } 
                
                \Begin{
                    
                    Encode deterministic relations in $\mathcal{M}_{cpd}$ into CNF\label{alg:AOBB-KstarMAP-MAIN:initialize-MiniSat}\;
                    
                    $\pi \leftarrow$ search path initialized with a dummy root node $r$ \label{alg:AOBB-KstarMAP-MAIN:initialize-dfs-to-dummy}\;

                    $H_{K^{*}} \leftarrow$ tables precomputed by $h_{K^*}(r)$

                    $H_{Z_{\gamma}} \leftarrow$ tables precomputed by $h_{Z_{\gamma}}(r)$ for each $\gamma$
    
                    \While{$EXPAND(\pi, \PT)$}{ \label{alg:AOBB-KstarMAP-MAIN:expand}
                        
                        \uIf{$ConstraintPropagation(\pi) = false$}{ \label{alg:AOBB-KstarMAP-MAIN:CPP}
                            $PRUNE(\pi)$ 
                        }
                    
                        \uElseIf{$\exists \gamma \in \varphi$ s.t. $ub_{Z_\gamma}(\pi,H_{Z_{\gamma}}) < S_{\gamma}$}{ \label{alg:AOBB-KstarMAP-MAIN:SSP}
                            $PRUNE(\pi)$ 
                        }
                        
                        \uElseIf{ $X \in \boldsymbol{R}$ }{ \label{alg:aobb-kstar:non-tip-node}
                            \If{ $ub_{K^*}(\pi, H_{K^{*}}) < lb_{K^*}$ }{ \label{alg:AOBB-KstarMAP-MAIN:UBP}
                                $PRUNE(\pi)$ 
                            }
                        }
                        
                        \While{$\pi$ \textnormal{has no unexpanded children}}{ \label{alg:AOBB-KstarMAP-MAIN:backtrack}
                                $BACKTRACK(\pi)$ 
                        }
                    }
    
                    \Return $lb_{K^*} = K^*MAP(\Mcpd)$
                }
                \end{footnotesize}
            \end{algorithm}

            \begin{algorithm}[h]
                \caption{\wMBEKstar{}}
                \label{alg:wMBE-KstarMAP}
                \begin{footnotesize}
                  \SetInd{0.25em}{0.55em}
                \Input{
                    CPD Graphical model $\mathcal{M}_{cpd}$  (Def \ref{def:cpd-graph-mod}); \: i-bound $i$;\\
                    constrained variable order $d = [X_1,...,X_n]$\\
                }
                \Output{
                    upper bound on the \KstarMAP: $ub_{K^{*}\!MAP}(\mathcal{M}_{cpd})$
                }
    
                \Begin{
                    Partition the functions $f \in \boldsymbol{F}$ into buckets $B_n,...,B_1$ s.t. each function is placed in the bucket corresponding to the highest-index variable in its scope.
                    
                    \ForEach{$k = n...1$}{ \label{alg:wMBE-KstarMAP:reverse-ordering}
                        Generate a mini-bucket partitioning of the bucket functions $\boldsymbol{MB_k} = \{ MB_k^{(1)},...,MB_k^{(T)} \}$ s.t. $|scope(f_{MB_{k}^{(t)}})| \leq i$, for all $MB_k^{(t)} \in \boldsymbol{MB_k}$ \label{alg:wMBE-KstarMAP:partition-mini-buckets}
                        
                        \uIf{$X_k \in \boldsymbol{MAP}$}{
                            \ForEach{$MB^{(t)}_k \in \boldsymbol{MB_k}$}{
                                $\lambda^{(t)}_k \leftarrow \max_{X_k} f_{MB_{k}^{(t)}}$ \label{alg:wMBE-KstarMAP:maximization}
                            }
                        }
                        \Else{
                            \uIf(\tcp*[f]{upper-bound for numerator}){$X_k \in \boldsymbol{C_B}$}{ 
                                \label{alg:wMBE-KstarMAP:BEGIN-upper-bound-bucket}
                                Select positive weights $\boldsymbol{w} = \{ w_1,...,w_T \}$ s.t. $\sum_{w_t \in \boldsymbol{w}} w_t = 1$
                                
                                \ForEach{$MB^{(t)}_k \in \boldsymbol{MB_k}$}{
                                    $\lambda^{(t)}_k \leftarrow \sum_{X_k}^{w_t} f_{MB_{k}^{(t)}}$ \label{alg:wMBE-KstarMAP:upper-bound-power-sum}
                                }
                                \label{alg:wMBE-KstarMAP:END-upper-bound-bucket}
                            }
                            \ElseIf(\tcp*[f]{lower-bound for denominator}){$X_k \in \boldsymbol{C_U}$}{
                                \label{alg:wMBE-KstarMAP:BEGIN-lower-bound-bucket}
                                Select a negative weight $w_1$ and positive weights $\boldsymbol{w} = \{ w_2,...,w_T \}$ s.t. $\sum_{w_t \in \boldsymbol{w}} w_t = 1$
                                
                                \ForEach{$MB^{(t)}_k \in \boldsymbol{MB_k}$}{
                                    $\lambda^{(t)}_k \leftarrow \sum_{X_k}^{w_t} f_{MB_{k}^{(t)}}$
                                    \label{alg:wMBE-KstarMAP:lower-bound-power-sum}
                                    
                                    \If{$scope(\lambda^{(t)}_k) \subseteq \R$}{
                                        $\lambda^{(t)}_k \leftarrow 1/\lambda^{(t)}_k$ \label{alg:wMBE-KstarMAP:invert-bucket-message}
                                    } 
                                }
                                \label{alg:wMBE-KstarMAP:END-lower-bound-bucket}
                            }
                        }
                        Add each $\lambda^t_k$ to the bucket of the highest-index variable in its scope.
                    }
                    \Return $\lambda_1 = ub_{K^{*}\!MAP}(\mathcal{M}_{cpd})$
                }
            \end{footnotesize}
            \end{algorithm}

    \section{Boosting \AOBBKstar} \label{sec:boosting}

        The original \AOBBKstar algorithm presented by \citet{pezeshki2022uai-aobb-for-cpd-2022} showed promise with good performance compared to state-of-the-art \BBKstar on small problems with rigid rotamers.  However it suffered from scalability issues with its performance decreasing on problems with three mutable residues and being unable to solve problems with four or more mutable residues.  In this paper we advance \AOBBKstar by presenting \AOBBKstar-b (boosted) with modifications to improve scalability.  These enhancements, which are outlined below, are a mix of CPD domain-specific enhancements as well as principled enhancements that can be generalized to other graphical models tasks and problem domains.

        \subsection{Boosted \wMBEKstar}
        
            As noted by the authors of \citet{pezeshki2022uai-aobb-for-cpd-2022}, a main cause of the scalability limitations of the \AOBBKstar was a sometimes weak or unbounded heuristic estimate by \wMBEKstar (Algorithm \ref{alg:wMBE-KstarMAP}).  This occurred primarily because of difficulties in the lower-bounding computations corresponding to the denominator of \Kstar and lead to loose upper bounds on the \KstarMAP, or - in the case of a zero-valued lower bound - an all-together unbounded \KstarMAP estimate. Such loose bounds (or lack of bounds) do not allow pruning during search, drastically enlarging the traversed search space. 

            To alleviate this, we introduce \textbf{\wMBEKstar-b (boosted)} with three sequential improvements to \wMBEKstar: (1) adjustment of the power-sum mechanism (lines \ref{alg:wMBE-KstarMAP:upper-bound-power-sum},\ref{alg:wMBE-KstarMAP:lower-bound-power-sum}) to produce non-zero lower-estimates at the cost of losing bound guarantees, (2) adjustment to the cost shifting mechanism to prevent cost-shifts with zeros, and (3) maximization with finite values over infinite ones (line \ref{alg:wMBE-KstarMAP:maximization}).  The specific adjustments and justifications for these modifications are explained next.

            \textbf{1) Enforcing non-zero lower estimates.}
                \wMBEKstar{}, uses a power-sum computation (lines \ref{alg:wMBE-KstarMAP:upper-bound-power-sum},\ref{alg:wMBE-KstarMAP:lower-bound-power-sum}) leveraging Holder's inequality to compute bounds (with a version using negative weights for lower bounding \citep{liu11-bounding-partition-function}). Deriving inspiration from \textit{\textbf{Domain-Partitioned MBE}} presented in \citet{pezeshki2022uai-aobb-for-cpd-2022} we adjust the computation to omit zeros in the lower-bounding power-sum (line \ref{alg:wMBE-KstarMAP:lower-bound-power-sum}) thus forcing non-zero estimates for consistent sub problems.  

                \begin{definition}[Zero-Omitted Weighted Function]
                    The zero-omitted $w$ weighted function of a function $f$ with scope $\Y$ is: $f^{\triangleleft w}(y) := f(y)^{w}$ for $f(y) \neq 0$ and $0$ otherwise.
                \end{definition}

                \begin{definition}[Zero-Omitted Power Sum]
                    The zero-omitted power sum of a function $f$ that includes $X$ in its scope is: $\sum_{X}^{\triangleleft w} f := (\sum_{X} f(x)^{\triangleleft \frac{1}{w}})^{w}$ where $\frac{0}{0} := 0$.
                \end{definition}
                
                Specifically, line \ref{alg:wMBE-KstarMAP:lower-bound-power-sum} is changed to: $\lambda^t_k \leftarrow \sum_{X_k}^{\triangleleft w} f_{MB_{k}^{(t)}}$.
                
                This modified version can no longer guarantee a lower bound, however boundedness can be retained if the omitted zeros correspond to conditions for bounded domain-partitioning \citep{pezeshki2022uai-aobb-for-cpd-2022}: 
                
                \begin{theorem}[Bounded Domain-Partitioning]
                    Consider three variables $X$, $Y$, and $Z$ and objective
                    \begin{align}
                        obj = \sum_{X} f(x,y) \cdot g(x,z)
                    \end{align}
                    Let $X' = \{ x \in X | g(x,z) \neq 0  \}$ be a set such that $\epsilon_{X'} =  min_{x \in X'} g(x,z)$. Since this makes $\epsilon_{X'} > 0$ we can derive: 
                    \begin{align}
                        obj =& \sum_{x \in X'} f(x,y) \!\cdot\! g(x,z) + \!\! \sum_{x \in X \setminus X'} f(x,y) \!\cdot\! g(x,z) \\
                        =& \sum_{x \in X'} f(x,y)  \!\cdot\! g(x,z) \;\;\geq\;\; \epsilon_{X'} \!\cdot\! \sum_{x \in X'} f(x,y)\\
                        \;>&\; 0
                    \end{align}
                    when $f(x,z)$ is not identically zero over $X'$.
                \end{theorem}
                
                which allows for weighted mini-buckets to retain (and improve) bounds.

            \textbf{2) Cost shifting with non-zero values.}
                We similarly adjust the cost-shifting mechanism (described in \cite{flerova2011-mbe-mm, IhlerFDO12}) restricting cost-shifts to be only with non-zero values. This again helps prevent numerical instabilities and ensures a positive lower bound for consistent sub problems.  This does not disrupt bound guarantees.

            \textbf{3) Maximizing finite values.}
                Due to the above-mentioned adjustments, the resulting \Kstar approximation for any [partial] configuration of the residues that are consistent will necessarily have a finite positive value (the upper bound estimates on the $K^* = \frac{Z_{B}}{Z_{U}}$ numerator are inherently finite and, now, the lower estimate of the denominator is also forced to be finite for consistent sub problems).
                Due to this result, during the maximization step (line \ref{alg:wMBE-KstarMAP:maximization}), we can now instead maximize only over the available finite values.

        \subsection{Tuning search}

            Two key enhancements were also made to \AOBBKstar search.

            \textbf{Prioritizing the wild-type assignment.}
                Unlike many other problem domains, when performing protein re-design a good initial assignment to the variables is known ahead of time: that which corresponds to the wild-type protein.  Thus, we force the wild-type to be explored first, ensuring that we initialize our search with a powerful lower bound.

            \textbf{Prioritizing nodes with a finite heuristic value.}
                Since \wMBEKstar-b produces infinite \Kstar estimates only for invalid configurations, we adjust node ordering during search to first explore nodes that have a finite heuristic value.  This ensures that consistent configurations are traversed first. 

        In the Empirical Evaluation section we evaluate the performance impact of these changes.

    \section{Dynamic Heuristics} \label{sec:dynamic-heuristics}

    Next we explore using dynamic heuristic re-computations to improve bounds and enhance pruning during search (ex. \citep{lam2014beyond}). \AOBBKstar-DH (Algorithm \ref{alg:aobb-kstar-dh-MAIN}) is a general framework for dynamic heuristic use with \AOBBKstar.

          \AOBBKstar-DH employs search similarly to \AOBBKstar with the exception that, at each node expansion not resulting immediately in pruning, there is a decision made whether or not to dynamically recompute a new \Kstar upper-bounding heuristic conditioned on the current search path (line \ref{alg:aobb-kstar-dh-MAIN:dh-recomputation-condition}).  This decision 
          is based on two hyper-parameters: $maxDepth$ (a maximum depth at which to consider recomputations) and $dhThreshold$ (a numerical bound on existing heuristic estimates over which re-computations occur). 
          These hyper-parameters serve to regulate the frequency of dynamic heuristic re-computations since they can be costly both in time and memory. (In particular, \wMBEKstar{} is exponential in its $i\mhyphen bound$ hyper-parameter).  When pruning or backtracking past the point of the most recent heuristic re-computation, the \Kstar heuristic tables $H_{K^{*}}$ are rolled back to cached tables from previous computations (not shown explicitly).
          Using an upper-bounding heuristic we have:

        \begin{algorithm}[t!]
            \caption{\AOBBKstar-DH}
            \label{alg:aobb-kstar-dh-MAIN}
            \begin{footnotesize}
                \SetInd{0.25em}{0.55em}
                \DontPrintSemicolon 
            \Input{
                CPD graphical model $\mathcal{M}_{cpd}$  (Def \ref{def:cpd-graph-mod}); \\
                pseudo tree $\PT$ guiding node expansions;\\
                $K^*$ upper-bounding heuristic function $h_{K^*}(.)$;\\
                $Z_\gamma$ upper-bounding heuristic function $h_{Z_\gamma}(.)$
            }
            \Output{
                $K^*MAP(\mathcal{M}_{cpd})$
            } 
            
            \Begin{
                
                Encode deterministic relations in $\mathcal{M}_{cpd}$ into CNF\label{alg:aobb-kstar-dh-MAIN:initialize-MiniSat}\;
                
                $\pi \leftarrow$ search path initialized with a dummy root node $r$ \label{alg:aobb-kstar-dh-MAIN:initialize-dfs-to-dummy}\;

                $H_{K^{*}} \leftarrow$ tables precomputed by $h_{K^*}(r)$

                $H_{Z_{\gamma}} \leftarrow$ tables precomputed by $h_{Z_{\gamma}}(r)$ for each $\gamma$

                \While{$EXPAND(\pi, \PT)$}{ \label{alg:aobb-kstar-dh-MAIN:expand}
                    
                    \uIf{$ConstraintPropagation(\pi) = false$}{ \label{alg:aobb-kstar-dh-MAIN:CPP}
                        $PRUNE(\pi)$ 
                    }
                
                    \uElseIf{$\exists \gamma \in \varphi$ s.t. $ub_{Z_\gamma}(\pi,H_{Z_{\gamma}}) < S_{\gamma}$}{ \label{alg:aobb-kstar-dh-MAIN:SSP}
                        $PRUNE(\pi)$ 
                    }
                    
                    \uElse{

                        \uIf{$depth(\pi) \leq maxDepth$ \textnormal{\textbf{and}} $H_{K^*}(\pi) > dhThreshold$}{ \label{alg:aobb-kstar-dh-MAIN:dh-recomputation-condition}
                            $H_{K^{*}} \leftarrow$ tables recomputed by $h_{K^*}(\pi)$
                        }
                        
                        \uIf{ $X \in \boldsymbol{R}$ }{ \label{alg:aobb-kstar-dh-MAIN:non-tip-node}
                            \If{ $ub_{K^*}(\pi, H_{K^{*}}) < lb_{K^*}$ }{ \label{alg:aobb-kstar-dh-MAIN:UBP}
                                $PRUNE(\pi)$ 
                            }
                        }
                        
                    }
                    
                    \While{$\pi$ \textnormal{has no unexpanded children}}{ \label{alg:AOBB-KstarMA-dh-MAIN:backtrack}
                            $BACKTRACK(\pi)$ 
                    }
                }

                \Return $lb_{K^*} = K^*MAP(\Mcpd)$
            }
            \end{footnotesize}
        \end{algorithm}

        \begin{theorem}[correctness, completeness]
            \AOBBKstar-DH is sound and complete, returning the optimal \Kstar value of a corresponding amino-acid configuration that does not violate the subunit-stability constraints.
        \end{theorem}

        \paragraph{maxDepth.}
            The $maxDepth$ parameter ensures that dynamic heuristics are not recomputed past a predetermined depth, bounding the maximum number of times re-computation can occur, as demonstrated next.

       \paragraph{Complexity of \AOBBKstar-DH \wMBEKstar Computations.}
            Since re-computation of the heuristic may occur anywhere in the search tree up to $maxDepth$ yielding exp($maxDepth$) number of nodes, and since each re-computation is $exp(i)$ given i-bound $i$, we can bound the time complexity as $O(k^{maxDepth + i})$, where $k$ is the maximum domain size encountered.  
            However, the number of nodes explored during search can be far smaller due to pruning, and a tighter heuristic can further reduce the explored space. Letting $\Lambda_{DH}$ represent the number of nodes explored with DH, the time complexity bound can be expressed as $O(\Lambda_{DH} \cdot k^{i}$). If we let $\Lambda$ be the number of nodes explored without $DH$,
            the time complexities with and without DH are $O(\Lambda_{DH} \cdot k^i)$ vs. $O(\Lambda)$ respectively.
            Thus, if $\frac{\Lambda}{\Lambda_{DH}} > k^{i}$ the use of DH will be cost-effective.
            Finally, since \AOBBKstar is a depth-first algorithm, the bound on the space overhead is linear in $maxDepth$ and $exp(i)$ for \wMBEKstar.
                
        \paragraph{dhThreshold.}
            Dynamic heuristic re-computations aim to improve bounds and enhance pruning. However, if the existing heuristic value is already tight, the cost of re-computation may outweigh traversing the search space with the current heuristic. Determining what is considered "already tight" can be uncertain for general search tasks. However, in the case of protein re-design valid solutions must have a cost similar to the wild-type. Therefore, initially, we can provide $dhThreshold$ values relative to the native wild-type \Kstar value, improving them as better solutions emerge.

    \section{UFO: A Principled Scheme to Introduce Determinism} \label{sec:ufo}

        It is well known that utilizing constraint propagation (CP) as a tool for pruning inconsistent search paths can greatly speed up search \cite{dechter-book-2ndEd,mateescu08-journal-mixed-networks,DBLP:books/daglib/0024906}.  Similar ideas have been explored in mixed integer programming \citep{danna2005exploring}. More recently in the scope of protein design \citet{pezeshki2022uai-aobb-for-cpd-2022} demonstrated that introducing artificially generated determinism by underflowing function values under a provided threshold (Definition \ref{def:fxn-underflow}) can further leverage CP and  enhance the speed of solving \Kstar optimization problems.  

        As our last set of algorithmic improvements
        we present 1) Algorithm \ref{alg:ufo}: UFO (underflow-threshold optimization) describing a general methodology for choosing underflow-thresholds
        (Definition \ref{def:fxn-underflow}) with certain characterizations, and 2)  \AOBBKstar-UFO, 
         \AOBBKstar augmented with a CPD-specific UFO scheme.

        \begin{definition}[$\tau$-underflow of $f$, $f_{\tau}$] \label{def:fxn-underflow}
            Let $f$ be a non-negative function and $\tau \in \mathbb{R}^{+}$. The $\tau$-underflow of $f$ is $f_{\tau}(x) = f(x)$ if $f(x) \geq \tau$ and $0$, otherwise. 
        \end{definition}
        
        \begin{definition}[$\tau$-underflow of $\mathcal{M}$, $\mathcal{M}_{\tau}$]
            For $\mathcal{M} \! = \! \langle\mathbf{X,D,F}\rangle$, the $\tau$-underflow of $\mathcal{M}$ is $\mathcal{M}_{\tau} = \langle\mathbf{X,D,F_{\tau}}\rangle$, where $\mathbf{F_{\tau}} = \{ f_{\tau} \;|\; f \in \boldsymbol{F}  \}$.
        \end{definition}

        \subsection{Underflow-threshold Choice}

            \begin{algorithm}[t!]

                \caption{UFO}
                \label{alg:ufo}
                \begin{footnotesize}
                    \SetInd{0.25em}{0.55em}
                    \DontPrintSemicolon 
                \Input{Graphical model $\mathcal{M} = \langle \X,\D,\F \rangle$; SAT solving algorithm, $SAT(.)$
                ; time limit for binary search; a deflation factor $0 < \delta \leq 1$}
                \Output{A proposed threshold $\tau$ to use}
                
                \Begin{
    
                    \uIf{$SAT(\mathcal{M}) = False$}{
                        return $FAILURE$
                    }
    
                    $\tau_{min} = 0$; \:
                    $\tau_{max} = \max_{\F,\X} f(\xx)$\\
                    $\tau = \frac{\tau_{max} + \tau_{min}}{2}$\\
    
                    \While{time remains for $\tau$ binary search}{ \label{alg:ufo-MAIN:BEGIN-t-binary-search}
                        \uIf{$SAT(\mathcal{M_{\tau}}) = False$}{ \label{alg:ufo-SAT-CHECK}
                            $\tau_{max} = \tau$ \\
                        }
                        \uElse{
                            $\tau_{min} = \tau$ \\
                        }
                        $\tau = \frac{\tau_{max} + \tau_{min}}{2}$ \\
                    }\label{alg:ufo-MAIN:END-t-binary-search}
    
                    $\tau = \tau_{min} \cdot \delta$ \label{alg:ufo:MAIN:deflate-threshold}\\

                    \Return $\tau$
                }
                \end{footnotesize}
            \end{algorithm}

            Clearly larger underflow-thresholds lead to more determinism and consequently more aggressive CP pruning. However, if the threshold is set too high, the resulting model becomes inaccurate and may even become inconsistent leaving no configuration capable of producing a non-zero value. 

            \begin{definition}[Inconsistent Model]
                A model is said to be inconsistent if $\; \forall \xx \in D_{\X}, \; \prod_{\F} f(x) = 0$.
            \end{definition}
            
            Therefore it is useful to find a threshold that is as high as possible yet still results in a consistent model. 
            To achieve this UFO employs binary search to find the largest threshold that still results in a satisfiable model (lines \ref{alg:ufo-MAIN:BEGIN-t-binary-search}-\ref{alg:ufo-MAIN:END-t-binary-search}).
            Then UFO decreases the threshold using a hyper-parameter $\delta$ (line 14) to enable a wider array of solutions.

            Note that UFO operates under the assumption that satisfiability of a model can be determined quickly. This is not true in general, nevertheless we have found that the satisfiability sub-task underlying many optimization problems tends to be easy. In other cases, satisfiability can be approximated by constraint propagation schemes
            \citep{dechter2003-constraints-book}.

        \paragraph{\AOBBKstar-UFO.}

            \AOBBKstar-UFO (Algorithm \ref{alg:ufo-aobb-kstar}) empowers \AOBBKstar by generating an underflowed model ${\mathcal{M}_{cpd}} _{\tau}$ with $\tau$ determined by the UFO scheme specially adjusted for CPD, denoted UFO\textsubscript{cpd}.  This modified UFO\textsubscript{cpd} performs underflows on the Boltzmann transformed $E^{sb}$ and $E^{pw}$ functions (see Equation  \ref{eq:boltzman-transformmed-energies-and-constraints}) and replaces the general satisfiability check in the UFO algorithm (Algorithm \ref{alg:ufo}, line \ref{alg:ufo-SAT-CHECK}) with one that enforces satisfiability of the wild-type sequence, thus ensuring a lower bound on the quality of solutions.

            \begin{algorithm}[t!]
                \caption{\AOBBKstar-UFO}
                \label{alg:ufo-aobb-kstar}
                    \SetInd{0.25em}{0.55em}
                    \DontPrintSemicolon 
                \Input{$\mathcal{M}_{cpd}$  (Def \ref{def:cpd-graph-mod}); $x_{wt}$, wild-type assignment to $X$; SAT solving algorithm, $SAT(.)$
                ; time limit for binary search; a deflation factor $0 < \delta \leq 1$; pseudo tree $\PT$ guiding search;
                $K^*$ upper-bounding heuristic function $h_{K^*}(.)$; $Z_\gamma$ upper-bounding heuristic function $h_{Z_\gamma}(.)$;}
                \Output{approximation to the true $K^{*}MAP(\mathcal{M})$}
                
                \Begin{
                    $\tau \leftarrow UFO_{cpd}(\Mcpd, x_{wt}, SAT(.), time\mhyphen limit, \delta)$\\
                    $K^{*}{}' \leftarrow AOBB{\text -}K^{*}({\Mcpd}_{\tau}, \mathcal{T}, h_{K^*}(.), h_{Z_\gamma}(.))$ \\
                    \Return $K^{*}{}'$
                }
            \end{algorithm}

            \paragraph{UFO for other graphical model tasks.} As UFO is a general scheme, it can be useful for a myriad of graphical model tasks.  A preliminary investigation into its use can be found in the report \href{https://www.ics.uci.edu/~dechter/publications.html}{Exploring UFO's}.

    \section{Empirical Evaluation} \label{sec:empirical-evaluation}

        \subsection{Experimental methodology}
            \paragraph{Benchmarks.}
                We performed empirical evaluation on benchmarks derived from re-design problems for real proteins provided by the Bruce Donald Lab at Duke University.  To gradually increase difficulty, small problems with two mutable residues (with five to ten total residues) were incrementally enlarged by making more of the residues mutable.  Experiments were performed on the "Expanded" problem set from \citet{pezeshki2022uai-aobb-for-cpd-2022} consisting of 12 problems with 3 mutable residues, a new set of 32 problems expanded to have 4 mutable residues, and a set of 18 problems expanded to have 5 mutable residues.  The names of the newly created benchmarks are shown with three parts: d[g]-[M]-[p] (eg. d27-4-1), where [g] represents the problem design number as obtained from the Donald Lab, [M] indicates the number of mutable residues after enlarging, and [p] is a single digit representing the specific permutation of the M residues that were made mutable. The resulting conformation spaces for these problems ranged from on the order of $10^{6}$ for 3 mutable residues to $10^{11}$ for 5 mutable residues.
            
            \paragraph{Algorithms.}
                We experimented with 5 algorithms: \AOBBKstar; \AOBBKstar-b (boosted) with an improved \wMBEKstar-b heuristic and search enhancements (Section \ref{sec:boosting}); \AOBBKstar-b-DH, \AOBBKstar-b with dynamic heuristics (Section \ref{sec:dynamic-heuristics}); \AOBBKstar-b-UFO, \AOBBKstar-b empowered with a CPD-specific UFO scheme (Section \ref{sec:ufo}); and \BBKstar, state-of-the-art best-first search algorithm in comprehensive CPD software OSPREY 3.0 \citep{ojewole18-bbkstar, hallen18-osprey-3}.
                
                Each \AOBBKstar derived algorithm was implemented in C++.  \AOBBKstar-b-DH dynamic heuristic re-computations were regulated with $maxDepth = 2$ and $dhThreshold = 10^{20} \cdot K^{*}_{wt}$, where $K^{*}_{wt}$ is the wild-type \Kstar value.  The UFO scheme used by \AOBBKstar-b-UFO performed binary search in log-space and decreased the resulting threshold with $\delta = 0.2$ (Algorithm \ref{alg:ufo}: UFO, line \ref{alg:ufo:MAIN:deflate-threshold}). Because the \AOBBKstar-b algorithms use the \wMBEKstar-b heuristic which does not guarantee bounds, they do not guarantee discovery of the optimal \Kstar (ie. they are not complete).  Similarly, schemes empowered with UFO lose optimality guarantees.
                
                \BBKstar is implemented in Java, was set to use rigid rotamers, and given a bound-tightness of $1\times10^{-200}$\textsuperscript[\footnote{\BBKstar's bound tightness parameter does not correlate directly with an $\omega$-approximation.  See \citet{ojewole18-bbkstar}.}\textsuperscript].  Despite the extremely small bound tightness parameter, \BBKstar still performs noticeably as an approximate algorithm.
                
                Experiments were run on a 2.66 GHz processor, and given 4 GB of memory and a time limit of 1hr for each problem.  As \BBKstar can take advantage of parallelism, it was given access to 4 CPU cores.
                
        \subsection{Results}
    
            \paragraph{Comparing \AOBBKstar-b vs \AOBBKstar.}
                In Table \ref{tbl:vanilla-comparison} we examine performance of \AOBBKstar-b (with tightened \wMBEKstar-b) vs. \AOBBKstar on the "Expanded" benchmark set from \citet{pezeshki2022uai-aobb-for-cpd-2022} with three mutable residues.  We compare solution quality and speed of the two algorithms.  The i-bound of \AOBBKstar-b was set to $i=4$. For \AOBBKstar we use the best performing i-bound as reported previously.  We highlight in blue any better \Kstar solutions and any significantly faster completion times (equal to or under 80\% of the competing algorithm's completion time).  The wild-type \Kstar value ("wt \Kstar\!") is also shown.
    
                The highlighted blue times show \AOBBKstar-b finishing significantly faster for half of the problems. It also solves a problem that \AOBBKstar could not.  Finally, \AOBBKstar-b is able to find the optimal solution for each of these problems (although it does not prove optimality).
                
                \begin{tablefigure}[!htb]
                    \centering
                	\caption{Performance of \AOBBKstar-b vs \AOBBKstar on benchmarks with 3 mutable residues from \citet{pezeshki2022uai-aobb-for-cpd-2022}. Displayed are the i-bound ("i") used by each, their respective best-found \Kstar value ("Soln"), their completion time ("Time"), and, as reference, the wild-type \Kstar value ("wt \Kstar").}
            		\includegraphics[scale=0.65]{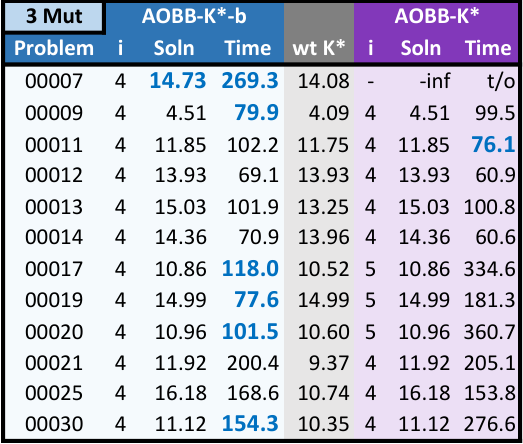}
            		\label{tbl:vanilla-comparison}
                \end{tablefigure}
                    
            \paragraph{Evaluating Dynamic Heuristics.}
                Table \ref{tbl:dynamic-heuristic-comparison-condensed} compares \AOBBKstar-b with and without the dynamic heuristic scheme described in Section \ref{sec:dynamic-heuristics} on problems with 3 or 4 mutable residues that both algorithms found optimal solutions for within an hour.  We compare the size of the explored search space between the two algorithms (counting the number of OR and AND nodes of the residue variables traversed) and highlight when there are differences.
    
                We see that dynamic heuristic re-computation reduces the size of the traversed search space in the majority of problems.  In two cases (highlighted in red) dynamic heuristics cause an increase in the search space.  This may occur when dynamic heuristic re-computation causes the \Kstar estimate for a node to increase (specifically by decreasing the denominator $Z_{U}$ estimate which \wMBEKstar-b does not guarantee to be a lower bound), preventing the node from being pruned.
        
                \begin{tablefigure}[!htb]
                    \centering
                	\caption{Comparison of the explored search space by \AOBBKstar-b with and without use of a dynamic heuristic.  Displayed are the respective i-bounds ("i") used, best-found \Kstar solutions ("Soln"), completion times ("Time"), and the size of the traversed AND/OR search space (number of residue OR and AND nodes).}
            		\includegraphics[scale=0.65]{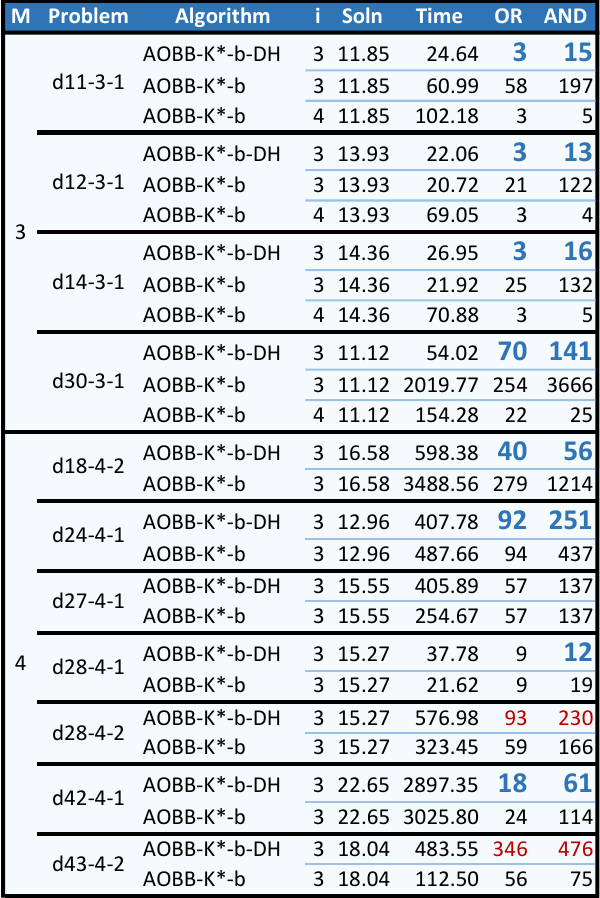}
            		\label{tbl:dynamic-heuristic-comparison-condensed}
                \end{tablefigure}
        
            \paragraph{UFO Impact and Cross Comparisons.}
                Table \ref{tbl:cumulative-comparison-condensed} compares the performance of \AOBBKstar-b-UFO, \AOBBKstar-b-DH, \AOBBKstar-b, and \BBKstar on problems with three, four, and five mutable residues.  The \AOBBKstar-based algorithms are displayed in a top-down ranking per problem, with the best ranking algorithm placed at the top. Ranking is based first on the quality of \Kstar found and then by the speed at which their respective solution was first discovered (measured in seconds and denoted "Anytime", highlighting the anytime nature of \AOBBKstar search).  Large text highlights the value responsible for the algorithm's higher ranking, and blue color indicates that \BBKstar was outperformed.
    
                From the rank-based ordering of the algorithms, the competitiveness of the UFO scheme is apparent.  The frequency of blue coloring shows the algorithms' competitiveness against \BBKstar on the problems with three and four mutable residues.  On problems having 5 mutable residues the \AOBBKstar-b schemes begin to struggle.  This is likely due to the loss of bounds from the \emph{boosted} modifications of \wMBEKstar-b in conjunction with a low i-bound, heavy underflows, and longer message passing for these larger problems.  Nevertheless, \AOBBKstar-UFO is still able to find good solutions, sometimes better than that of \BBKstar.
    
                We also see the potential of the \AOBBKstar-DH scheme in terms of run-time in Table \ref{tbl:cumulative-comparison-condensed}.
                On many problems shown, \AOBBKstar-b-DH performs better than \AOBBKstar-b (sometimes due to a better solution found, other times due to finding good solutions faster).  However, \AOBBKstar-b-DH's performance with respect to \AOBBKstar-b is less homogeneous when including the easier problems from Table \ref{tbl:dynamic-heuristic-comparison-condensed} which were omitted from Table \ref{tbl:cumulative-comparison-condensed}.
    
                Finally, although \AOBBKstar-b generally ranked lower than the other \AOBBKstar-b variants, it keeps up with \BBKstar through problems with 4 mutable residues (previously out of range for \AOBBKstar), and even finds respectable solutions for some 5-mutable-residue problems.
    
                \begin{tablefigure}[!htb]
                    \centering
                	\caption{Comparison of the \AOBBKstar-b-[UFO/DH] schemes and \BBKstar on problems ranging from 3 to 5 mutable residues.  Shown is the i-bound used, best-found \Kstar solution (recomputed without underflow-thresholding), the time at which the best-found solution was first discovered ("Anytime"), and the completion time ("Time").  For reference, the wild-type \Kstar solution is also shown.}
            		\includegraphics[scale=0.62]{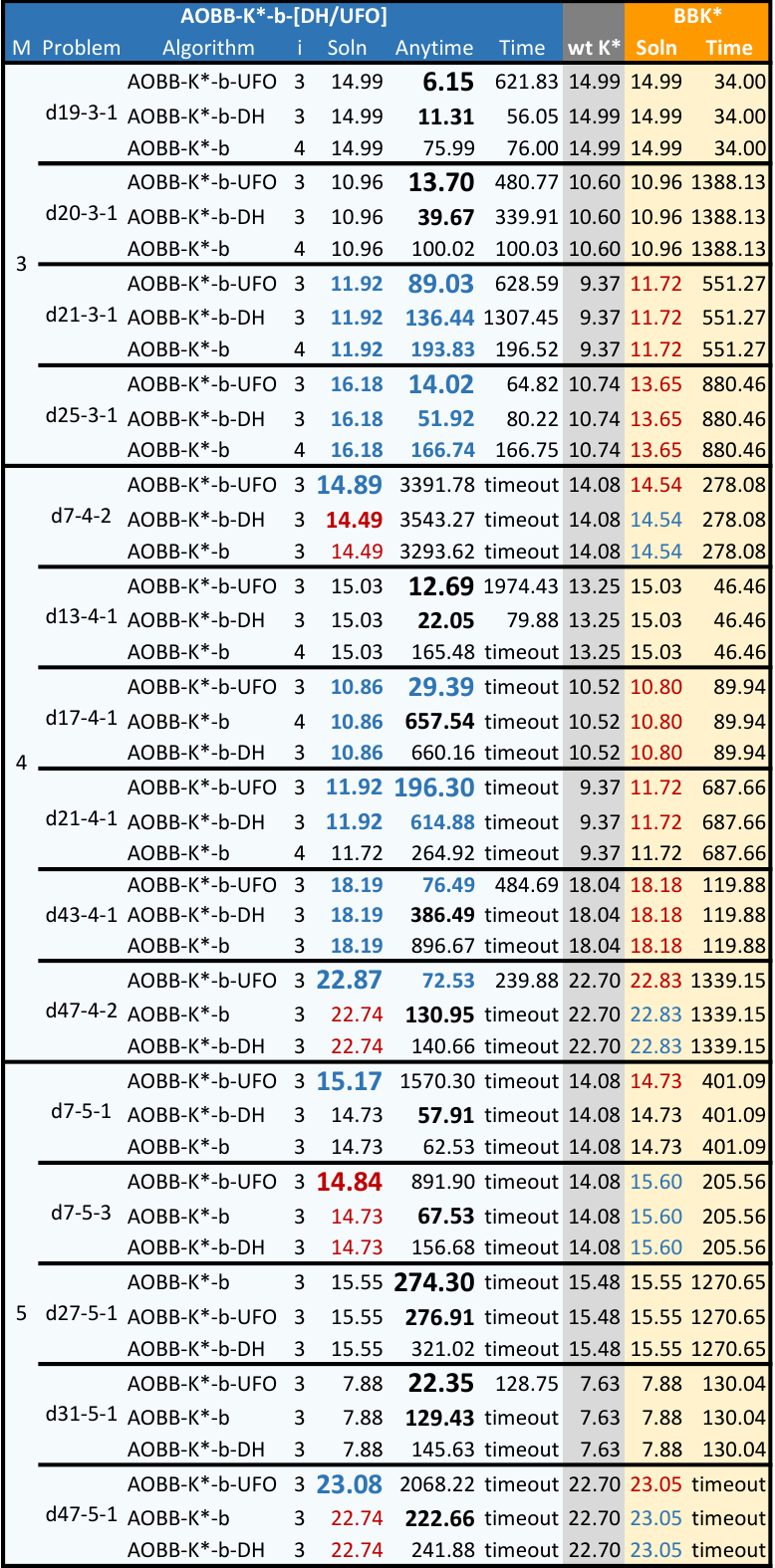}
            		\label{tbl:cumulative-comparison-condensed}
                \end{tablefigure}

            \textbf{Determinism.}
                A major factor that can lead to decreased performance on large problem for \AOBBKstar schemes (eg. problem d7-5-3) is that domain sizes increase significantly with an increasing number of mutable residues.  This  restricts \wMBEKstar{} to using lower i-bounds and reduces its accuracy. For example, \AOBBKstar could use an i-bound of 4 for problems with 3 mutable residues, but was restricted to an i-bound of 3 for problems with 5 mutable residues in Table \ref{tbl:cumulative-comparison-condensed}.  To explore the potential of moving to more compact representations that could enable use of higher i-bounds, determinism in  \wMBEKstar-b's computed messages for \AOBBKstar-b-UFO were evaluated.  For problems with 5 mutable residues, the largest tables generated by \wMBEKstar-b often had a determinism ratio of > 0.95 - namely 95\% of the entries were zeros.  This insight adds motivation to moving to other representations that can take advantage of the repeated determinism, such as relational representations.

            {\bf Summary of Empirical Results.}
                \AOBBKstar-b, \AOBBKstar-b-DH, and \AOBBKstar-b-UFO have now achieved scalability to problems with 5 mutable residues. The UFO scheme demonstrated strong performance in particular, competitive with \BBKstar for these problems. Analysis of \AOBBKstar-b-DH's explored search space shows its promise, but the current naive implementation showed limited performance on larger problems. Lastly, analysis of \wMBEKstar-b in the presence of UFO revealed a high level of determinism indicating that more compact representations may be beneficial.  (Additional results can be found in the \href{https://www.ics.uci.edu/~dechter/publications.html}{Supplemental Materials}).

    \section{Conclusion and Future Directions}

        \textbf{Conclusion.} This work introduced several improvements to the protein re-design algorithm \AOBBKstar to enhance its scalability. Refinements to its \wMBEKstar heuristic were presented sacrificing bound guarantees for tighter estimates, and adjustments were made to its node value-ordering strategy during search. These enhancements were implemented in \AOBBKstar-b (boosted) which scaled up to problems with 5 mutable residues whereas \AOBBKstar could only produce solutions to problems with 3 mutable residues. To further enhance pruning during search, the dynamic heuristic scheme \AOBBKstar-DH was introduced.  Evaluation with a naive implementation showed the promise of dynamic heuristics being incorporated into \AOBBKstar. Additionally, UFO - a new underflow-thresholding scheme for introducing artificial determinism to strengthen constraint propagation - was introduced. A specialized version of UFO for CPD was incorporated into \AOBBKstar-b as \AOBBKstar-b-UFO and showed competitive performance against state-of-the-art \BBKstar on problems of up to 5 mutable residues. Evaluation of these algorithms was done using 62 real-protein benchmarks involving three to five mutable residues.
            
        \textbf{Future Directions.}
            We leave the following as future directions: to
            \textbf{1.} investigate modifications to \wMBEKstar-b that mitigate the risk of violating boundedness;
            \textbf{2.} adapt richer implementations of dynamic heuristics \citep{lam2014beyond};
            \textbf{3.} consider look-ahead schemes \citep{DBLP:journals/jair/LamKLD17};
            \textbf{4.} extend to finding the n-best \Kstar's such as approaches by \cite{DBLP:journals/jair/FlerovaMD16, ruffini-diversity-and-optimality-cpd-cfns};
            \textbf{5.} investigate the trade-offs of the dynamic heuristic hyper-parameters;
            \textbf{6.} explore UFO utility in other graphical model tasks;
            \textbf{7.} explore more compact representations of \wMBEKstar that can take advantage of the high levels of determinism \citep{mateescu08-journal-mixed-networks,larkin03-bayesian-inference-in-the-presence-of-determinism} or other scalable heuristics \citep{lee16-exact-to-anytime-MMAP};
            \textbf{8.} apply the proposed approaches to real biological tasks;
            \textbf{9.} extend these algorithms by incorporating other state-of-the-art inference schemes, especially approximate schemes \citep{yanover2002approximate,toulbar2-2016,lou2018anytimespace,lou2018finite,Marinescu19-any-recursive-bf-search-for-bounding-mmap, marinescu18-stochastic-anytime-mmap, marinescu18-jair-AO-search-mmap}.

        \subsubsection*{Acknowledgements} 
            Thank you to reviewers of this paper for their valuable comments and suggestions. We also acknowledge use of ChatGPT, an AI language model developed by OpenAI, in refining text. This work was supported in part by NSF grant IIS-2008516.

\clearpage

    \bibliography{ref}

\end{document}